\documentclass[twocolumn]{aastex62}
\usepackage{xcolor}
\usepackage{natbib}
\usepackage{hyperref}

\hypersetup{
    pdfnewwindow=true,      
    colorlinks=true,        
    linkcolor=blue,         
    citecolor=black,        
    filecolor=blue,         
    urlcolor=blue           
}

\begin{document}
\def\gsim{ \lower .75ex \hbox{$\sim$} \llap{\raise .27ex \hbox{$>$}} }
\def\lsim{ \lower .75ex\hbox{$\sim$} \llap{\raise .27ex \hbox{$<$}} }
\def\clov#1{{\bf[#1 -- Clovis]}}

\shorttitle{Constraining the Fraction of CC-SN Harboring Choked Jets with HE Neutrinos}
\shortauthors{Guetta, Rahin, Bartos, Della Valle}
 
\title{Constraining the Fraction of Core-Collapse Supernovae Harboring Choked Jets with High-energy Neutrinos}


\author{Dafne Guetta}
\email{dafneguetta@braude.ac.il}
\affiliation{ORT Braude, Karmiel, Israel}
\author{Roi Rahin}
\email{roirahin@gmail.com}
\affiliation{Technion, Haifa, Israel}
\author{Imre Bartos}
\email{imrebartos@ufl.edu}
\affiliation{Department of Physics, University of Florida, PO Box 118440, Gainesville, FL 32611-8440, USA}
\author{Massimo Della Valle}
\email{massimo.dellavalle@inaf.it}
\affiliation{Capodimonte Observatory, INAF-Naples , Salita Moiariello 16, 80131-Naples, Italy}
\affiliation{European Southern Observatory, Karl-Schwarzschild-Straβe 2, D-85748 Garching bei
München, Germany}

\begin{abstract}
The joint observation of core-collapse supernovae with gamma-ray bursts shows that jets can be launched in the aftermath of stellar core collapse, likely by a newly formed black hole that accretes matter from the star. Such gamma-ray bursts have only been observed accompanying Type Ibc supernovae, indicating a stellar progenitor that lost its Hydrogen envelope before collapse. According to recent hypothesis it is possible that jets are launched in core-collapse events even  when the progenitors still retain their Hydrogen envelopes, however, such jets are not able to burrow through the star and will be stalled  into the interior of the progenitor star before escaping. These jets are called chocked jets.  High-energy neutrinos produced by such choked jets could escape the stellar envelope and could be observed. Here we examine how multi-messenger searches for high-energy neutrinos and core-collapse supernovae can detect or limit the fraction of stellar collapses that produce jets. We find that a high fraction of jet production is already limited by previous observational campaigns. We explore possibilities with future observations using LSST, IceCube and Km3NET.
\end{abstract}


\section{Introduction}

The mechanism driving core-collapse supernova (CCSN) explosions is still unclear. A common explanation is neutrino driven outflows \citep{Colgate82}, however a recent idea is that jets could play an important role in SN explosions \citep{Piran2019}. CCSNe associated with long duration gamma-ray bursts (GRBs), sometimes called “Hypernovae”, provide a clear evidence for strongly aspherical explosions \citep{Maeda2008}, suggesting a jet-like activity. 
Observations strongly suggest that a tiny fraction of CC (GRB-SNe) are powered by jets (see \cite{GDV2007}, \cite{Soderberg2006}, \cite{Berger2003}). Recently we have observed in GRB171205A/SN 2017iuk the associated “cocoons” \cite{Izzo2019}. This fact opens the interesting possibility that also broad-lines SNe without GRBs - which are energetically similar to GRB-SNe ($\sim10^{52}$ erg) may be powered by “cocoons”.


\cite{Piran2019} explored that a similar mechanism might be active also in "standard" CCSNe types not associated with GRBs. Such a scenario is suggested by observations indicating energy deposition by so called "choked" jets in a cocoon within the stellar envelope. 

\cite{Soker2011} discusses a scenario characterized by no-relativistic jets in CC-SNe.
The cocoon eventually breaks out from the star releasing energetic material at very high, yet sub-relativistic, velocities.  It is possible to identify this fast moving material by looking at the very early time supernova spectra, as this component has a unique signature. Recently, a clear evidence has been provided indicating such a deposition by \citep{Izzo2019}, who detected spectroscopic signatures of iron group elements moving at $\sim $\,120,000\,km\,s$^{-1}$ originating from the innermost parts of the exploding progenitor star of SN\,2017iuk. However, whether or not CCSNe that do not harbor GRBs might be powered by jet/cocoons activity is still a matter of active discussion within the community. The most natural interpretation of the energetic fast moving component observed in the early spectra of the supernovae by Piran et al. 2019, is that this is the cocoon’s matter that breaks out from the progenitor. In a few cases such as SN\,2008D/GRB\,080109 \citep{Mazzali2008}, SN\,2006aj/GRB\,060218  \citep{Campana2007} and GRB 171205A/2017iuk \citep{Izzo2019} we might have even seen the direct thermal emission of this hot cocoon material. In this case Piran's interpretation implies the existence of powerful jets within these supernovae. 
These jets have difficult to penetrate the progenitor, therefore these objects are simply observed as energetic supernovae or hypernovae in the optical band and do not have accompanying gamma-ray emission. The jets that do not succeed to drill the envelope are labelled "chocked" jets (CJ).

It may be that a significant fraction of core-collapse SNe, and possibly all those that have not lost their Hydrogen envelope, harbor choked jets. The shocks involved in these hidden jets may be the source of high energy neutrinos observed by IceCube (see i.e. \citep{Senno2016} and \citep{He2018}). 


In chocked jets neutrinos and gamma-rays are produced by the decay of pions produced in the interaction of accelerated protons and thermal photons.  While these sources are transparent to neutrinos they are opaque to gamma rays photons as these particles are 
produced inside the stellar envelope. This can justify the lack of association between observed GRBs and IceCube neutrinos.
Therefore these sources are dark in GeV-TeV gamma-rays, and do not contribute  to
the Fermi diffuse gamma-ray background implying that the Waxman-Bahcall bound and the Fermi constrain on the diffuse gamma-ray emission do not apply to these sources \citep{Murase2016}.

 These characteristics have been used by  several authors, e.g. \citep{Murase2013,Murase2016,Senno2016,Senno2017,Denton2017, Esmaili2018}, to show that chocked jets – so called failed GRBs may be the  possible sources for the observed diffuse neutrino flux.

In this paper we check if we can use the neutrino emission to test the hypothesis that most supernovae produce jets (e.g., \citealt{Piran2019}). If only a small faction of CCSNe is powered by jets (e.g. only GRB-SNe) the number of neutrinos revealed by neutrino detectors will be smaller than the number expected if the totality of CCSNe (including, type II, Ibc and GRB-SNe) are powered by jets.

We investigate the detectability of high-energy neutrinos from choked jets within CCSNe, and the limits set by a non-detection on the fraction of CCSNe that harbor jets. For this analysis we assume that  the conditions set by \cite{Senno2016} are satisfied and use the model of \cite{He2018} to calculate the expected neutrino flux from a single source. In their estimate they include the relevant microphysical processes such as multi-pion production in pp and p-gamma interactions, as well as the energy losses of mesons and muons.

 

\section{Neutrino emission estimate from chocked jets}

 We consider the possibility that a fraction of rapidly rotating massive stars at the end of their lives undergo core collapse, form a compact star or a black hole, and produce a jet that it is stalled before it breaks through the star \citep{Meszaros2001Jet, Macfadyen2001} .
We take the stellar parameters given in \citep{Meszaros2001Jet}.

Following \citep{Senno2016} we assume that the photons are free to move inside the jet  as the plasma is optically thin inside the jet but cannot escape as the envelope or circumstellar medium outside is largely optically thick to Thomson scattering. Since the photons are trapped inside the jet the protons 
can interact with these thermal photons very efficiently.
We can assume that the fraction of protons converted in pions, the so-called $f_{\pi}$, is almost 1 in this process. The radiation constraints apply to shocks in the envelope material as well as those in the choked jet.  With their model Senno et al. 2016 have estimated the diffuse neutrino flux from chocked jets. In this paper we are interested in estimating the flux from a single source.

We consider the possibility that electrons and protons are accelerated in the internal shock model and
estimate the neutrino flux and spectrum  from a single source following the model of 
\cite{He2018}.  The model of \cite{He2018} is valid if the conditions described in \cite{Senno2016} are met:   the jet is choked, protons are accelerated to high energies efficiently, and thermal photons are produced in the jet head and propagate into the internal shock region.
 In this model the accelerated protons interact with the photons from the choked jet head and produce pions. The pions decay into high energy neutrinos. \cite{He2018} estimate the neutrino spectra numerically, taking into account micro-physical processes.  There is a low energy cutoff in the neutrino spectra
 due to the threshold of the photo-meson interaction. The other cutoff is at high energy due to the photo-meson cooling of protons, or the synchrotron cooling of pions and muons.
 
  In this model The parameters that affect most the neutrino flux are the Lorentz factor, the duration of the jet and the luminosity of the jet.  In our analysis we use the same values for these parameters taken in \cite{He2018}.  These are $\Gamma =100, 10$, $t=100, 3.3\times10^{4}$\,s and $L=10^{51}, 3\times 10^{48}$\,erg\,sec$^{-1}$. \cite{He2018} show how different set of parameters imply different neutrino fluxes and spectra  see Figure 3 of \cite{He2018}. Their results are consistent with  the diffuse flux found by \cite{Senno2016} and therefore with the upper limits given from IceCube on point source analysis and on diffuse flux analysis.

\section{Neutrino detection}

High-energy neutrinos interact with nucleons present in the detector producing secondary particles, which travel faster than the speed of light in the sea or ice, therefore induce the emission of Cherenkov light. 

Currently operating high-energy neutrino detectors include, IceCube is a cubic-kilometer observatory located at the geographic South Pole \citep{detIce}; the ANTARES detector deep in the Mediterranean sea \citep{detAnt}; and KM3Net \citep{detKm3}, also in the Mediterranean, currently under construction.

We determine the total number of expected neutrinos by a source by integrating the neutrino fluence convoluted with the effective energy of the detector over the energy range 1-100 TeV:
\begin{equation}
N(r)=\int_{1\,TeV}^{100\,TeV}  \;\frac{dN_{\nu}(E(1+z))}{dE_{\nu}}A(E_{\nu},\delta)dE_{\nu} 
\label{eq:Nr}
\end{equation}
where $r$ is the luminosity distance, $\frac{dN_{\nu}}{dE_{\nu}}$ is the neutrino spectral fluence, $A(E_{\nu},\delta)$ is the effective area of the neutrino detector, as a function of the neutrino energy $E_{\nu}$ and of the source declination.
 In our analysis we consider the IceCube effective area for time-dependent searches using muon neutrinos from point sources \citep{detIce3}.
 The area for Point Sources is found using the results of \cite{detIce4} and the link https://icecube.wisc.edu/science/data/PS-3years.
 
  The number of neutrinos expected from a source at distance of $\sim$ 1 Gpc is consistent with the upper limits found by IceCube as shown in \cite{He2018}.

\section{Fraction of supernova observations with detected neutrino counterpart}

We now estimate the fraction $f_{\nu}$ of CCSNe detected electromagnetically that will also be detected via high-energy neutrinos. We convolve $N(r)$ with the frequency of occurrence of CCSNe as a function of redshift. Due to the shortness of the lifetime of the progenitors of core-collapse supernovae, typically a few million years, we can use star formation history as proxy of the history of the CCSN rate (e.g. \citealt{Hopkins2006,Madau2014}). We find that the expected fraction of supernovae within distance $r_{\rm max}$ that will have a detected astrophysical high-energy neutrino counterpart is
\begin{equation}
    f_{\nu} = \frac{f_{\rm jet}f_{\rm b}\int_{0}^{r_{\rm max}}(1 - \rm{Poiss}(0, N(r))) \rho(r) 4\pi r^2 dr}{\int_{0}^{r_{\rm max}} \rho(r) 4\pi r^2 dr}
\label{eq:fnu}    
\end{equation}
where $f_{\rm b}$ is the jet beaming factor, defined as the fraction of the sky in which the jet emits high-energy neutrinos, $f_{\rm jet}$ is the fraction of CCSNe that produce jets, Poiss$(k,\lambda)$ is the Poisson probability of measuring $k$ for average value $\lambda$, and $r_{\rm max}$ is the maximum luminosity distance out to which the CCSN can be detected optically. This fraction depends on the neutrino detection threshold through $N(r)$ defined in Eq. \ref{eq:Nr}. In this definition we do not include the possibility that a background neutrino can be coincident by chance with the supernova, which we separately account for below.

Here we consider $r_{\rm max}$ for two representative facilities for ongoing and future observations: the Zwicky Transient Facility (ZTF; \citealt{RauchICZTF}) and the Large Synoptic Survey Telescope (LSST; https://www.lsst.org/scientists/scibook). For ZTF, we assume a magnitude limit of $R\sim 19$ for a typical 30\,s exposure \citep{Bellm2017,RauchICZTF}. For LSST we assume a magnitude limit of R$\sim$24 for a typical exposure of $2\times 15$\,s \citep{LSST2019}. After assuming an absolute magnitude at maximum for CCSNe of is M= -17.5 \citep{patat94}), we find that they can be detected out to $r\sim200$\,Mpc with ZTF, and out to $r\sim1.5$\,Gpc with LSST.  The cosmological effects are taken into account in the analysis. For the LSST case, the redshift effects become important. 

With these observational limits, using Eq. \ref{eq:fnu}  and the model parameters (Model 1) $\Gamma=100$, $t=100$ sec and $L=10^{51}$erg/sec we find that a fraction $f_{\rm \nu}^{\rm ZTF} \approx 0.5f_{\rm b}$ of CCSNe detected by ZTF might produce neutrinos detectable with IceCube or Km3NET. For LSST, this fraction becomes $f_{\rm \nu}^{\rm LSST} \approx 0.01f_{\rm b}$. In order to check the effects of the model parameters on our results we consider another set of model parameters (Model 2) like $\Gamma=100$, $t=3.3\times 10^{4}$ sec and $L=3.3 \times 10^{48}$erg/sec, we find that a fraction $f_{\rm \nu}^{\rm ZTF} \approx 0.03f_{\rm b}$ of CCSNe detected by ZTF might produce neutrinos detectable with IceCube or Km3NET. For LSST, this fraction becomes $f_{\rm \nu}^{\rm LSST} \approx 5\times 10^{-4}f_{\rm b}$. For another set of parameters (Model 3) like $\Gamma=10$, $t=3.3\times 10^{4}$ sec and $L=3.3 \times 10^{48}$erg/sec,  we find that a fraction $f_{\rm \nu}^{\rm ZTF} \approx 0.01f_{\rm b}$ of CCSNe detected by ZTF might produce neutrinos detectable with IceCube or Km3NET. For LSST, this fraction becomes $f_{\rm \nu}^{\rm LSST} \approx 10^{-4}f_{\rm b}$.

The same estimate can be performed for SNe-Ibc which are a fraction of CC, about 30\% \citep{Botti2017} but, on average,  slightly brighter (up to M$\approx -18$ ) so  detectable with LSST up to 2.3 Gpc and with ZTF to 280Mpc. 

We find that,  for the model parameters of Model1 a fraction $f_{\rm \nu}^{\rm ZTF} \approx 0.2f_{\rm b}$ of SNe-Ibc detected by ZTF might have IceCube or Km3NET neutrinos detections. For SNe-Ibc detected by LSST, this fraction is $f_{\rm \nu}^{\rm LSST} \approx 4\times 10^{-3}f_{\rm b}$. For the other set of parameters, Model2, we find  a fraction $f_{\rm \nu}^{\rm ZTF} \approx 0.02f_{\rm b}$ of SNe-Ibc detected by ZTF might have IceCube or Km3NET neutrinos detections. For SNe-Ibc detected by LSST, this fraction is $f_{\rm \nu}^{\rm LSST} \approx 4\times 2\times10^{-4}f_{\rm b}$. In the case of Model 3 we find  a fraction $f_{\rm \nu}^{\rm ZTF} \approx 0.007f_{\rm b}$ of SNe-Ibc detected by ZTF might have IceCube or Km3NET neutrinos detections. For SNe-Ibc detected by LSST, this fraction is $f_{\rm \nu}^{\rm LSST} \approx 4\times 6.2\times10^{-5}f_{\rm b}$ 

Finally we consider Hypernovae (without GRBs) which are about 7\% of SNe-Ibc \citep{GDV2007}. These sources are of particular interest because possibly they are the most suitable candidates to harbor chocked jets. HNe are as bright as -19 , so detectable up to 3 Gpc (to 400 Mpc with ZTF).

We find that for Model 1 a fraction $f_{\rm \nu}^{\rm ZTF} \approx 0.02f_{\rm b}$ of Hypernovae detected by ZTF will may produce neutrinos detectable by IceCube or Km3NET. For Hypernovae detected by LSST, this fraction is $f_{\rm \nu}^{\rm LSST} \approx 7\times 10^{-4}f_{\rm b}$.  For Model 2 we find that a fraction $f_{\rm \nu}^{\rm ZTF} \approx 0.008f_{\rm b}$ of Hypernovae detected by ZTF will may produce neutrinos detectable by IceCube or Km3NET. For Hypernovae detected by LSST, this fraction is $f_{\rm \nu}^{\rm LSST} \approx 9\times 10^{-5}f_{\rm b}$. For Model 3 we find that a fraction $f_{\rm \nu}^{\rm ZTF} \approx 0.003f_{\rm b}$ of Hypernovae detected by ZTF will may produce neutrinos detectable by IceCube or Km3NET. For Hypernovae detected by LSST, this fraction is $f_{\rm \nu}^{\rm LSST} \approx 3\times 10^{-5}f_{\rm b}$.

All these factors are smaller by a factor about 10 for ANTARES. 

The fractions of CCSNe producing neutrinos detectable with  IceCube or Km3NET, decrease from ZTF to LSST because the latter telescope will be able to discover an increasing number of supernovae out to very large distances therefore hardly recognizable by neutrinos detectors.  




\section{Neutrino background rate}

The main component for the background is the flux of atmospheric neutrinos, which is caused by the interaction of cosmic rays, high energy protons and nuclei, with the Earth's atmosphere. Decay of charged pions and kaons produced in cosmic ray interactions generates the flux of atmospheric neutrinos and muons. Their energy spectrum is about one power steeper than the spectrum of the parent cosmic rays at Earth, due to the energy dependent competition between meson decay and interaction in the atmosphere.

Supernovae have evolutionary time scale of the order of dozens of days. However, we expect that a jet will be driven and neutrinos will be emitted from SNe only for a time frame comparable to the duration of long GRBs, i.e. about a minute. The time of core collapse is much more uncertain than this time frame, with characteristic uncertainties of hours-days. This means that the relevant time frame for background accumulation is hours-days. Here, we consider 1 day as a characteristic uncertainty. Given that IceCube detects about 100,000 neutrino candidates per year \citep{detIce2}, and assuming 1\,deg$^2$ directional uncertainty, the expected number of background neutrinos coincident with a given CCSN is $N_{\rm atm}\sim 10^{-2}$.



\section{How many supernova follow-ups will lead to a multi-messenger detection?}

We first consider a search for neutrinos coincident with a population of $N_{\rm sn}$ detected CCSNe. We want to know how many supernovae will lead to a $3\sigma$ detection of neutrinos, assuming that a fraction $f_{\rm jet}$ of the supernovae produce jets and therefore high-energy neutrinos with luminosities estimated above. The expected number of background neutrinos candidates detected in coincidence with the $N_{\rm sn}$ supernovae is $10^{-2}N_{\rm sn}$. The expected number of signal neutrinos from these supernovae is $N_{\rm sn}f_{\nu}$. To discover a connection at $3\sigma$ significance level, we require the expected number of signal neutrinos to be equal to $3\sigma$ uncertainty of the number of background neutrinos, i.e. the following equation needs to be satisfied:
\begin{equation}
N_{\rm sn}f_{\nu} = 3\sqrt{10^{-2}N_{\rm sn}}.
\end{equation}
To obtain quantitative results for the different models considered here, we adopt a beaming factor $f_{\rm b}\approx 0.1$ \citep{2007ApJ...662.1111L}. 

The obtained numbers of supernovae of different types required for detection at $3\sigma$ significance level are listed in Table \ref{table:SNnumber}. 
We see that, for our CCSN model with rate density $7\times10^{-5}$\,Mpc$^{-1}$yr$^{-1}$, the required number of CCSN detections can be achieved in less than a year by either ZTF or LSST if the fraction of CCSNe producing jets is $\sim1$.

\begin{table}[h!]
\begin{center}
\begin{tabular}{ c|c|c}
  & ZTF & LSST \\ 
  \hline
 CCSN & 40 & $10^{5}$ \\  
 SN-Ibc & 200 & $6\times10^{5}$ \\
 Hypernova & $2\times 10^4$ & $2\times 10^7$ \\
\end{tabular}
\end{center}
\caption{Expected number of electromagnetically detected that need to be followed up with IceCube or KM3NeT to identify them as sources of high-energy neutrinos at $3\sigma$ significance level for different source types (column \#1), considering sources electromagnetically identified by ZTF (column \#2) or LSST (column \#3). For models SN-Ibc and Hypernova, we adopted source parameters from our Model 1. We assumed that a fraction $f_{\rm jet}=1$ of the sources produce jets (and therefore neutrinos) with beaming factor $f_{\rm b}\approx 0.1$. For other parameters, all results in this table scale as $f_{\rm jet}^{-2}f_{\rm b}^{-2}$.}
\label{table:SNnumber}
\end{table}

\section{How many neutrino follow-ups will lead to a multi-messenger detection?}

Let $f_{\rm SN}$ denote the currently unknown fraction of astrophysical neutrinos detected by IceCube / KM3Net that were produced by SNe. Given that the density of CCSNe follows the cosmic star formation rate, we can calculate the fraction of neutrinos from SNe that come from within a distance threshold. For ZTF's 200\,Mpc this fraction is 2\%, while for LSST's 1500\,Mpc it is 15\% \citep{2017PhRvD..96b3003B}. Therefore, the fraction of detected astrophysical neutrinos that originate from SNe detectable by ZTF is $0.02f_{\rm SN}$, while for LSST it is $0.15f_{\rm SN}$. Assuming optimistically that a significant fraction of astrophysical neutrinos come from SN jets, i.e. $f_{\rm SN}\sim1$, we find that about $50$ (6) neutrinos need to be followed up to find a coincident CCSN with ZTF (LSST). This number is proportionally higher if not all neutrinos originate from CCSNe. 

After assuming a temporal coincidence time window of 1 day and neutrino localization uncertainty of $1^\circ$, the probability of chance coincidence between a neutrino and a CCSN is $5\times10^{-4}$ for ZTF and 0.2 for LSST. For ZTF, following up about 100 neutrinos is expected to lead to 2 coincident CCSNe, which would be a statistically significant deviation ($3\sigma$) from the background. For LSST, following up 80 neutrinos would lead to an expected statistically significant ($3\sigma$) excess of signal neutrinos over the background. 

\section{Detection with neutrino multiplets}

 Detecting a single or even a double neutrino in coincidence with a SN is not by itself sufficient for discovery. A higher number of neutrinos coincident with a single SN would be needed to establish an astrophysical origin. 
Approximating the number fraction of SNe that result in the detection of three or more astrophysical neutrinos as $f_{\rm \nu}^3$, we find that the inverse of this fraction is significantly larger than the number of SNe needed for detection with a single neutrino (see Table \ref{table:SNnumber}), Therefore, we conclude that the follow-up of neutrino multiplets is not competitive with the follow-up of multiple supernovae with a single detected neutrino.

\section{Conclusion}

We investigated the prospects of probing jet production by a large fraction of core-collapse supernovae below the stellar envelope, as proposed by \cite{Piran2019}. We calculated the expected high-energy neutrino flux from choked jets below the stellar envelope and calculated their detectability with IceCube, Km3NET and ANTARES. We computed the number of follow-ups that need to be carried out in order to find coincident supernovae+neutrinos and obtain the fraction of supernovae that produce jets. We find that our results depend on the adopted parameters, such as Lorentz factor, burst duration and luminosity. We have explored the set of parameters of Model 1, which we consider to be the most realistic case. Our conclusions are the following:
\begin{itemize}
\item If all CCSNe produce jets with a beaming factor of 0.1 then we need about 40 CCSNe detected by a ZTF-like telescope, or $10^5$ CCSNe detected by LSST, to be able to establish neutrino emission and therefore jets from CCSNe. This can be achieved within 1 year of observations with either ZTF or LSST. If not all CCSNe produce jets then these numbers are proportionally higher. 
\item Considering the electromagnetic follow-up of astrophysical neutrinos, about 100 astrophysical neutrinos need to be followed up by a ZTF-like telescope or LSST in order to establish a statistically significant excess of temporally and directionally coincident CCSNe, according to our neutrino emission model. The required number of follow-ups is greater due to the fraction of high-energy neutrinos that are not astrophysical, and if not all CCSNe drive jets.
\item Searching for neutrino doublets coincident with CCSNe can lead to discovery even with a single such association. We find that the search for such doublets will lead to discovery on a similar time frame as the accumulation of singlets.
\item 
The results show that the search for distant Hypernovae will add a lot of noise and few neutrinos, therefore if we include Hypernovae it will take many years to make a detection.  We recommend a search strategy that does not look beyond CCSN distances in order to optimize signal to noise ratio.

\end{itemize}
 The results found in this paper are consistent with the ones already cited in this paper \cite{Senno2016,Senno2017,He2018} that estimate the diffuse neutrino flux obtained from choked jets and use the observed IceCube neutrino flux to constrain the neutrino production mechanism model.
Our results are also consistent with the ones found in a recent work of
\cite{Esmaili2018}, who searched for temporal and spatial coincidences between high-energy neutrino events in six years of IceCube data and SNe Ib/c that occurred in the same time period. They did not find any significant correlation. However they used their analysis to put a constrain on the total neutrino energy and therefore on the isotropic-equivalent energy of cosmic rays. Their results are consistent with the constraints found by \cite{Senno2017} who used one year sample of muon neutrinos in IceCube. Another interesting work on a similar topic is the one of \cite{Denton2017}, who perform a detailed analysis of the internal shock model for the production of high energy neutrinos and use the the IceCube results to constrain the model parameters. Our results on the effect of the model parameters are consistent with their conclusions.

\bibliographystyle{aasjournal}
\bibliography{Refs}





	









 

\end{document}